\documentclass[amsmath,amssymb,nofootinbib,notitlepage,superscriptaddress,twocolumn]{revtex4-2}

\usepackage{graphicx,mathtools,setspace,subfigure}
\usepackage[margin=1in,letterpaper]{geometry}
\usepackage[colorlinks=true]{hyperref}
\hypersetup{citecolor=blue}

\usepackage[bb=dsserif]{mathalpha}
\usepackage{bm}

\newtagform{brackets}{[}{]}
\usepackage{amsfonts}
\usepackage{color}
\usepackage{amsthm}
\usepackage{mathrsfs}
\usepackage{comment}
\usepackage{braket}
\usepackage{ulem}
\usepackage[export]{adjustbox}

\usepackage{tabularray}

\newcommand{\pd}{\partial}

\begin{document}

\title{Can black holes evaporate past extremality?}
\author{Samuel E. Gralla}
\affiliation{Department of Physics, University of Arizona, Tucson, AZ 85721, USA}

\begin{abstract}
Black holes with sufficiently large initial charge and mass will Hawking-evaporate towards the extremal limit.  The emission slows as the temperature approaches zero, but still reaches the point where a single Hawking quantum would make the object superextremal, removing the horizon.  We take this semiclassical prediction at face value and ask: When the emission occurs, what is revealed?  Using a model of thin-shell collapse with subsequent accretion/evaporation by a null flux of ingoing positive/negative energy (charged Vaidya spacetime glued to a flat interior), we find that the matter re-emerges as a null shell that expands to infinity.  This expanding remnant has been bathed in the ingoing Hawking quanta during evaporation and presumably carries correlations with the outgoing quanta, offering the attractive possibility of studying information paradox issues in a setup where spacetime curvatures are globally small, so that quantum gravity is not required.  Even for ordinary black holes that evaporate down to the Planck size, we propose a radical new scenario for the interior: rather than forming a singularity, the collapsing matter settles onto an \textit{outgoing} null trajectory \textit{inside} the horizon for the entirety of evaporation.
\end{abstract}

\maketitle

\section{Introduction}

In the usual story of black hole evaporation, a black hole forms from collapse and then slowly shrinks due to Hawking radiation until it reaches Planck size, where the semiclassical approximation breaks down and its fate remains a mystery \cite{Hawking:1974rv,Hawking:1975vcx,Hawking:1976ra,Wald:1975kc,Unruh:1976db,Unruh:2017uaw,Almheiri:2020cfm,Witten:2024upt}.  This scenario is valid for the known types of astrophysical black holes (if isolated), which lose mostly angular momentum at first, and then lose their mass \cite{Page:1976df,Page:1976ki}.  However, the lack of massless charged particles changes the story for charged black holes with sufficient mass \cite{Gibbons:1975kk,Hiscock:1990ex}.  Once the temperature drops below the mass of the lightest charged particle, the Hawking effect can no longer efficiently produce charged particles, and the small electric field at the horizon also suppresses the Schwinger effect.

The discharge is thus exponentially suppressed at large black hole mass, so that Hawking radiation evolves a charged black hole towards its extremal (zero temperature) limit, with the mass $M$ approaching the charge $|Q|$ from above.  The Hawking radiation slows as the limit is approached, but the suppression is only by a power law, which cannot compete with the exponential suppression of discharge (see Ref.~\cite{Page:2000dk} for timescale estimates).  We therefore reach the point where the energy deviation from extremality becomes of order the Hawking temperature $T_H$,
\begin{align}\label{deviant}
    M-|Q| \approx T_H \approx \frac{\hbar^2}{2\pi^2 |Q|^3},
\end{align}
and the emission of a single quantum would naively make the spacetime super-extremal.

This has been interpreted as indicating breakdown of semiclassical physics \cite{Preskill:1991tb,Maldacena:1998uz,Almheiri:2016fws,Brown:2024ajk,Kraus:2025efu}, but we think by contrast that the semiclassical approximation should be excellent in such a low-curvature regime.  An underlying statistical/thermodynamic interpretation may become challenging, but the Hawking calculation does not invoke these concepts; it simply predicts the emission of quanta with a thermal spectrum.  We see no reason to doubt this prediction.  The energy \eqref{deviant} of a typical quantum is tiny and has a tiny effect on the geometry, so Hawking's fixed-spacetime assumption appears to be as well-satisfied as ever.  We will therefore accept evaporation past extremality as a \textit{prediction} of semiclassical physics, with a ``last quantum'' removing the horizon.  But then we are faced with a question: What is left behind?

\begin{figure}
    \includegraphics[width=\linewidth]{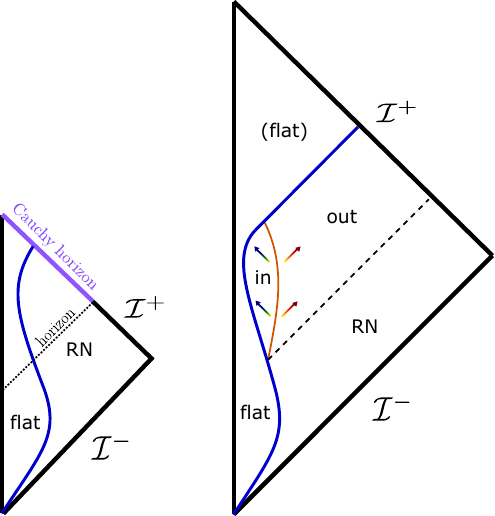}
    \caption{Penrose diagrams for charged spherical collapse, with and without backreaction.  The shell trajectory is shown as a blue line.  In Einstein-Maxwell theory (no backreaction---left panel), the spacetime is flat inside the shell and RN outside the shell; both an event horizon and a Cauchy horizon are formed.  In the Vaidya model of backreaction with evaporation past extremality (right panel), the Hawking radiation is modeled by regions of ingoing (``in'') and outgoing (``out'') charged Vaidya spacetime separated by a timelike pair formation front (brown curve) lying somewhere outside the shrinking apparent horizon.  No event horizon or Cauchy horizon forms in this model; the shell becomes null and escapes to infinity after evaporation past extremality.}\label{fig:new-penrose}
\end{figure}

The remnant's mass, spin, and charge must equal those of the macroscopic black hole at the end of evaporation, but beyond that it is hard to guess.  Further progress requires studying the matter that formed  the black hole, as it is this matter---perhaps dramatically transformed---that ultimately produces the remnant.  A baseline expectation can be set by studying spherical collapse in electrovacuum, where the generalized Birkhoff theorem forces the exterior to be the Reissner-Nordstr\"{o}m (RN) spacetime.  At least for a thin-shell model of the matter, one finds that the black interior does not contain a singularity, but instead contains a macroscopic shell forming a Cauchy horizon  \cite{de1967gravitational,Boulware:1973tlq,proszynski1983thin} (see Fig.~\ref{fig:new-penrose} left).

However, the Cauchy horizon is not expected to be stable to perturbations (classical or quantum-mechanical), since it receives highly blueshifted signals from the entire future of the exterior \cite{Simpson:1973ua}.  In essence, its existence and properties depend sensitively on everything that ever happens outside the black hole.  Even if there are no classical perturbations, the effects of the quantum vacuum---colloquially, the ingoing Hawking radiation---will accumulate at the Cauchy horizon.  The question of what remains after evaporation past extremality depends on the backreaction of the quantum fields on the spacetime metric.

While a complete calculation is out of reach at present, we still expect Einstein's equation to describe the bulk dynamics, since the spherical shell and Cauchy horizon remain macroscopic.  To make progress, we will adopt a simple classical model of this problem, where the collapsing matter is a charged shell, and the effects of Hawking radiation
are mocked up with null dust \cite{Hiscock:1980ze,Hiscock:1981xb}. 

 More precisely, we glue a flat interior to an ingoing charged Vaidya \cite{Bonnor:1970zz}  exterior at a matching radius corresponding to the shell position.  The exterior has a prescribed mass $M(v)$ that varies with advanced time $v$.  The increasing-mass case $M'(v)>0$ corresponds to energy falling into the black hole, while the decreasing case ($M'(v)<0$) models evaporation, with negative-energy null dust representing ingoing Hawking quanta.  In the latter case it is natural to also match to an outgoing Vaidya spacetime at a ``pair formation front'' outside the apparent horizon, but we do not perform the construction explicitly since it is identical to the original work \cite{Hiscock:1981xb}.

This model yields coupled ordinary differential equations for the shell radius $r$ and its material mass $m$, which we solve using a mixture of numerical and analytical techniques.  We find that the shell radius approaches the inner horizon radius $r_-=M-\sqrt{M^2-Q^2}$ during initial collapse, and remains near this (now time-evolving) radius during subsequent accretion or evaporation.  On the other hand, the shell mass $m$ is dramatically affected by the incoming radiation.  It grows exponentially during any initial period of accretion ($M'(v)>0$), a potential embarrassment of the model that we wish away by regarding $m$ as an unmeasurable bookkeeping parameter.  However, once evaporation takes over ($M'(v)<0$), then $m$ decreases to zero at a \textit{finite} time $v=\hat{v}$, after which the shell becomes null and outgoing.  When the horizon disappears after evaporation past extremality and the shell can finally escape, it simply expands at the speed of light. 

The model thus predicts that the remnant is an \textit{expanding null shell}.  As measured from infinity, the shell energy is slightly below the charge (making the spacetime slightly superextremal), 
\begin{align}
E=|Q|-\epsilon,
\end{align}
where $\epsilon \sim \hbar^2/(2\pi^2|Q|^3)$ according to \eqref{deviant}.  This outcome is illustrated in the right panel of Fig.~\ref{fig:new-penrose}.

The scenario of a macroscopic expanding remnant has implications for the information paradox.  This remnant has been bathed in the ingoing Hawking quanta of the entire evaporation process, so it should contain correlations with the outgoing Hawking quanta.  At least for formation and immediate evaporation (where the shell internal energy remains small), the spacetime curvatures are arbitrarily small everywhere throughout the entire evaporation process, such that we expect some kind of semiclassical approximation (with backreaction) to be valid.  One could imagine formulating and solving the relevant backreacted equations and explicitly accounting for the flow of information during a complete scenario of black hole formation and disappearance, without the need for quantum gravity.  A full understanding of this ``baby information paradox'' would surely provide new insight into the total-evaporation version.  We discuss these issues more fully at the conclusion of the paper.

Although we have worked in spherical symmetry, these results have potentially important implications for generic black holes.  The interior structure of a generic Kerr-Newman black hole closely resembles that of the RN black hole, with a Cauchy horizon and a timelike singularity.  Although more difficult to study than the spherically symmetric case, it seems very likely that collapse will still produce a macroscopic object forming a Cauchy horizon (covering up the timelike singularity).  We can then expect the evaporation to proceed as in our simple model.  We therefore suggest that realistic black holes---those in our universe---do not have singular interiors.  Rather, the interior consists of \textit{outgoing} matter (formerly collapsing or infalling), which loses mass, charge, and angular momentum as evaporation proceeds.  It would be very interesting to determine the fate of this matter after full evaporation.

This paper is organized as follows.  In Sec.~\ref{sec:model} we set up the equations of the model.  In Sec.~\ref{sec:collapse-without-backreaction} we discuss collapse without backreaction and in Sec.~\ref{sec:collapse-with-backreaction} we include backreaction.  In  Sec.~\ref{sec:limitations} we discuss limitations of the model, and in Sec.~\ref{sec:implications} we discuss its potential implications.  We use units with $G=c=4\pi \epsilon_0=k_B=1$ and adopt the metric signature $(-+++)$.

\section{Thin-shell model}\label{sec:model}

Charged thin-shell collapse in RN spacetime was studied in Refs.~\cite{de1967gravitational,Boulware:1973tlq,farrugia1979third,proszynski1983thin,Hawkins:1994sq}, and null dust has been used to model black hole  evaporation in Refs.~\cite{Hiscock:1980ze,Hiscock:1981xb} and many later references, with the charged case covered in Refs.~\cite{Kaminaga:1988pg,Strominger:1993yf,Jacobson:1997ge}.  We put the two ideas together, using the ingoing charged Vaidya spacetime outside of a thin shell.  The outgoing Vaidya region can be easily added on in an identical manner to the construction of \cite{Hiscock:1981xb}, so we do not consider it explicitly.

We use a plus/minus to denote the exterior/interior of the shell.  The metric inside the shell is flat, 
\begin{align}
    ds_-^2 = -dt^2 + dr^2 + r^2 d\Omega^2,
\end{align}
while the exterior is the ingoing charged Vaidya spacetime,
\begin{align}\label{gout}
    ds_+^2 = - f(r,v) dv^2 + 2 dv dr + r^2 d\Omega^2,
\end{align}
where
\begin{align}\label{f}
    f(r,v) = 1 - \frac{2M(v)}{r} + \frac{Q(v)^2}{r^2}.
\end{align}
When $|Q| < M$, the roots $r_{\pm}=M\pm\sqrt{M^2-Q^2}$ are called the inner and outer horizons.  These radii evolve as the mass and charge change.

We use the areal radius $r$ consistently.  The shell worldsurface $\Sigma$ is parameterized as
\begin{align}
    r=R(\tau), \quad v=V(\tau), \quad t=T(\tau).
\end{align}
Our intrinsic coordinates for the shell will be $x^a=(\tau,\theta,\phi)$.  We take $\tau$ to be the proper time of radial streamlines,
\begin{align}
    u^a = (1,0,0),
\end{align}
so that
\begin{align}\label{hab}
    ds^2_{\Sigma} = - d\tau^2 + R(\tau)^2 d\Omega^2.
\end{align}
The induced metrics computed from the left and the right are
\begin{align}
ds^2_{\Sigma_-} & = (- \dot{T}^2 + \dot{R}^2) d\tau^2 + R^2 d\Omega^2 \\
ds^2_{\Sigma_+} & = (-f \dot{V}^2 + 2 \dot{V}\dot{R}) d\tau^2 + R^2d\Omega^2,
\end{align}
where an overdot indicates the $\tau$ derivative.  To agree with \eqref{hab} we have
\begin{align}\label{unorm}
    -\dot{T}^2 + \dot{R}^2 = -f \dot{V}^2 + 2 \dot{V}\dot{R} = -1,
\end{align}
which are just the normalization conditions for the four-velocity of the shell as described in the interior and exterior,
\begin{align}
   u^{\mu_-} = (\dot{T},\dot{R},0,0), \quad u^{\mu_+} = (\dot{V},\dot{R},0,0),
\end{align}
We place a $\pm$ on coordinate indices to distinguish between the external coordinates $\mu_+=(v,r,\theta,\phi)$ and the internal coordinates $\mu_-=(t,r,\theta,\phi)$.

We may solve Eqs.~\eqref{unorm} to determine explicit expressions for $\dot{T}$ and $\dot{V}$,
\begin{align}
    \dot{T} &= \sqrt{1+\dot{R}^2}, \label{Tdot} \\
    \dot{V} &= \frac{1}{f}\left(\dot{R} +s \sqrt{f+\dot{R}^2}\right), \label{Vdot}
\end{align}
where we choose $\dot{T}>0$ to keep the correct time orientation, and
\begin{align}\label{sformula}
    s & = \textrm{sign} \left(f \dot{V}-\dot{R}\right).
\end{align}
We also require $\dot{V}>0$ to have the correct time orientation.  Enumerating the various cases, we see that in $f>0$ regions, we must have $s=+1$, but either sign of $\dot{R}$ is allowed, whereas in $f<0$ regions, both $s=\pm 1$ are allowed, while $\dot{R}$ must be negative.  The restriction on the sign of $\dot{R}$ is the usual statement that $f<0$ is a trapped region, where the entire light cone involves motion to smaller size.  The restrictions on $s$ can be understood by noting that $f\dot{V}-\dot{R}=-g_{\mu_+ \nu_+} (\partial_v)^{\mu_+} u^{\nu_+}$, so that $s$ is the sign of the ``Killing energy'' of a shell element.  (We use this terminology even though $\partial_v$ is a strict Killing field only in the special case of time-independent $M$ and $Q$.)  This explains why $s$  must be positive where $\partial_v$ is timelike ($f>0$), and can take either sign where $\partial_v$ is spacelike ($f<0$).

\begin{figure}[t]
    \includegraphics[width=.9\linewidth]{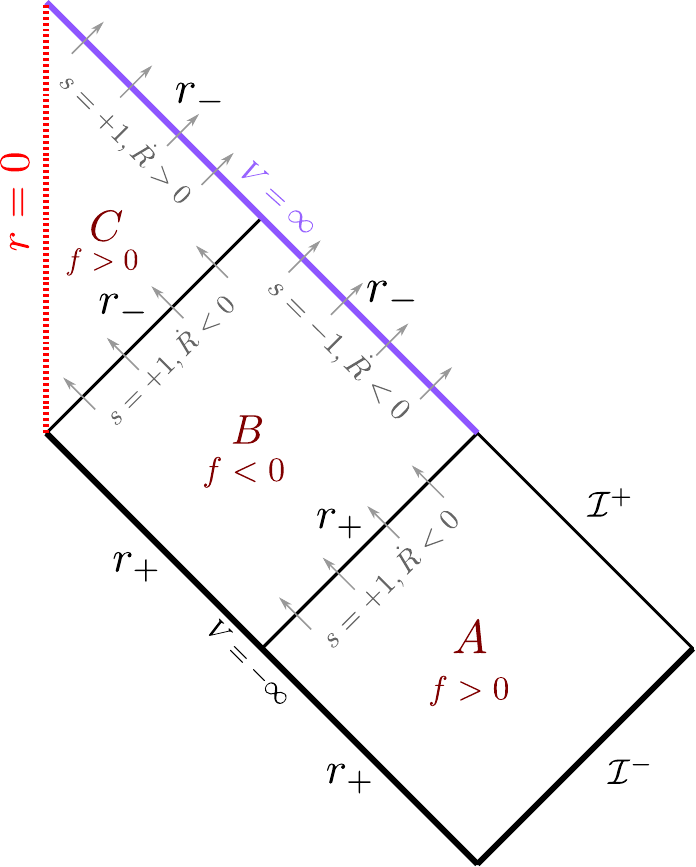}\caption{Penrose diagram illustrating the meaning of various definitions and coordinates.  The diagram precisely reflects the causal structure of the coordinate patch only for constant subextremal $M$ and $Q$ (RN spacetime), but it still forms good intuition for the slowly-evolving case we consider.  We use capital letters for the coordinates on the diagram, since mostly we will think about the history of a shell described by $v=V(\tau)$ and $r=R(\tau)$.  The gray equations indicate properties of these functions as $f=0$ horizons are crossed (indicated with gray arrows).  The regions between the horizons are labeled $A$, $B$, and $C$. The sign $s$ is defined in Eq.~\eqref{Vdot}.}\label{fig:regions}
\end{figure}

Notice that the behavior of $\dot{V}$ as $f \to 0$ depends critically on the signs of $\dot{R}$ and $s$.  If they have opposite sign then $\dot{V}$ smoothly approaches the value $-1/(2\dot{R})$ required by \eqref{unorm} at $f=0$ with finite $\dot{V}$, whereas if they have the same sign then $\dot{V}$ diverges as $2|\dot{R}/f|$.  In the former case the shell is crossing a finite-$v$ horizon (either region $A \to B$ or $B\to C$ in Fig.~\ref{fig:regions}), and in the latter case one is crossing the Cauchy horizon, either from $B$ or from $C$.

The normal vector to the surface may be expressed in internal or external coordinates as
\begin{align}
    n_{\mu_-}=(-\dot{R},\dot{T},0,0), \quad  n_{\mu_+}=(-\dot{R},\dot{V},0,0).
\end{align}
These expressions may be verified by confirming unit normalization and orthogonality to $u^{\mu_\pm}$.  The tangent vectors to the shell worldsurface are
\begin{align}
e^{\mu_\pm}_{(\tau)} & = u^{\mu_\pm} \\
e^{\mu_\pm}_{(A)} & = \delta^{\mu_\pm}{\!}_A,
\end{align}
where $A$ represents $\theta$ or $\phi$. The extrinsic curvatures are defined as
\begin{align}
    k^\pm_{ab} = \nabla_{\mu_\pm} n_{\nu_\pm} e^{\mu_\pm}_{(a)} e^{\nu_\pm}_{(b)}.
\end{align}
After repeated use of \eqref{unorm} and its $\tau$-derivative, we find
\begin{align}
k^-_{\tau \tau} & = \frac{-\ddot{R}}{\dot{T}} \\
    k^+_{\tau \tau} & = \frac{1}{1+ f\dot{V}^2}\left(-2 \ddot{R} \dot{V} + \dot{V}^3 \pd_v f - \dot{V} \pd_r f\right) \\
    k^-_{AB}  & = R \dot{T} \Omega_{AB} \\
    k^+_{AB} & = R(f \dot{V} - \dot{R} )\Omega_{AB},
\end{align}
with $\Omega_{AB}=\textrm{diag}(1,\sin^2 \theta)$ the sphere metric.  We will only need the spherical components $k^\pm_{AB}$, but for completeness we also include the expressions for $k^\pm_{\tau\tau}$. 

The shell stress tensor is given by \cite{Poisson:2009pwt}
\begin{align}
    - 8 \pi S_{ab} = (k^+_{ab}-k^-_{ab}) - ( k^+-k^-) h_{ab},
\end{align}
where $h_{ab}$ is the induced metric \eqref{hab} and $k^\pm=h^{ab} k^\pm_{ab}$ denotes the trace.  The four-dimensional stress-energy tensor is then
\begin{align}\label{T4d}
    T^{\mu_\pm \nu_\pm} = S^{ab} e^{\mu_\pm}_{(a)} e^{\nu_\pm}_{(b)} \delta(\ell),
\end{align}
where $\ell$ is a scalar field vanishing on $\Sigma$ such that $n_\alpha = \pd_\alpha \ell$.  In terms of the coordinate path $R(v)$ of the shell, we have $\delta(\ell) = \sqrt{f-2R'(v)}\delta(r-R(v)).$

The component $S_{\tau \tau}$ is often denoted $\sigma$ to reflect its interpretation as a proper energy per unit area.  This component is determined from $k^{\pm}_{AB}$ and the induced metric $h_{ab}$, and is equal to
\begin{align}\label{Stautau}
    S_{\tau \tau} = \sigma = \frac{1}{4\pi R}\left( \dot{T} + \dot{R} - f \dot{V} \right).
\end{align}
We also define the total energy $m$ as
\begin{align}
    m = 4\pi R^2 \sigma,
\end{align}
so that
\begin{align}\label{m}
    m = R(\dot{T} + \dot{R} - f \dot{V}).
\end{align}
We use the symbol $m$ following previous work, but this model cannot distinguish between rest mass and other forms of internal energy.  In general, $m$ should be interpreted as the total internal energy.  We will not need the formula for $S_{AB}$.

From \eqref{m} and \eqref{Tdot}--\eqref{sformula}, we find 
\begin{align}\label{Rdoteqn}
\sqrt{1+\dot{R}^2} = \frac{m^2-Q^2+2MR}{2mR}.
\end{align}
together with
\begin{align}\label{s}
    s = \textrm{sign}\left(\frac{-m^2-Q^2+2MR}{2m}\right).
\end{align}
Evidently, the sign $s$ flips at a ``branching radius''
\begin{align}\label{rb}
    r_b = \frac{Q^2+m^2}{2M}.
\end{align}
For $m>0$, Eq.~\eqref{s} may be written
\begin{align}\label{s2}
    s=\textrm{sign}(R-r_b).
\end{align}
As discussed below Eq.~\eqref{sformula}, both signs are allowed when $f<0$, but $s=+1$ is required for $f>0$.  Thus for $m>0$ we have a physical restriction
\begin{align}\label{branchrestriction}
    R>r_b \qquad \text{when } f>0.
\end{align}

Eq.~\eqref{Rdoteqn} gives the radius of a radial turning point $\dot{R}=0$ as
\begin{align}\label{r0}
    r_0=\frac{Q^2-m^2}{2(M-m)}.
\end{align}
Assuming the physical case $M>0$,  $m>0$, and $R>0$,  real solutions exist for $\dot{R}$ when
\begin{align}
    R & \leq r_0, \qquad \textrm{if } m>M \label{rangebound} \\
    R & \geq r_0, \qquad \textrm{if }  m<M. \label{rangeunbound}
\end{align}
These ranges hold instantaneously for the time-evolving parameters $m, Q, M$.   In the massless limit the range is
\begin{align}\label{r0massless}
    R \geq \frac{Q^2}{2M} \qquad (m \to 0).
\end{align}
This massless turning point is inside the inner horizon.

We may square \eqref{Rdoteqn} to yield 
\begin{align}\label{Rdot2}
    \dot{R}^2 = \left( \frac{m^2-Q^2+2MR}{2mR} \right)^2 - 1.
\end{align}
As discussed below Eq.~\eqref{sformula}, one must choose the negative root in the $f<0$ region, whereas both the positive and negative roots are allowed in $f>0$ regions.

In analyzing these equations, one must be careful to discard spurious solutions.  The principle is that one must be able to reconstruct 
a timelike, future-directed four-velocity $u^{\mu_+}=(\dot{V},\dot{R},0,0)$, or equivalently find a real solution for $\dot{R}$ together with a real, \textit{positive} solution for $\dot{V}$.  This amounts to ensuring that $\dot{R}$ and $s$ meet the sign restrictions discussed below Eq.~\eqref{sformula}. (We also require $f+\dot{R}^2 \geq 0$ so that the square root in \eqref{Vdot} yields a real value, but this property in fact holds automatically from the formula \eqref{Rdot2} for $\dot{R}^2$.)  In an $f>0$ region, $R$ must satisfy \eqref{branchrestriction} and \eqref{rangebound}--\eqref{rangeunbound}, and either the positive or negative branch of $\dot{R}$ may be chosen.  In the $f<0$ region, $R$ must satisfy \eqref{rangebound}--\eqref{rangeunbound}, $\dot{R}$ must be negative, and $s$ is determined from \eqref{s}.  These requirements are sufficient to generate a physical solution.

Eq.~\eqref{Rdoteqn} can be suggestively rearranged as
\begin{align}\label{energy}
    M = m \sqrt{1+\dot{R}^2} - \frac{m^2-Q^2}{2R},
\end{align}
which has an energy interpretation.  If we regard $M$ as the total shell energy, then \eqref{energy} divides it into material and field contributions.  Indeed, the first term is just $m\dot{T}$ and hence equals the  material energy according to observers in the flat interior region, while the second term is exactly equal to the field energy of a massive charged shell in Newtonian gravity.
  
We now impose conservation of stress-energy, as required by Einstein's equation.  In terms of quantities intrinsic to the shell worldvolume $\Sigma$, the condition is \cite{de1967gravitational}
\begin{align}\label{force}
    D^b S_{ab} = -F_a,
\end{align}
where $D$ is the covariant derivative on $\Sigma$ and
\begin{align}
    F_a = T_{\mu^+ \nu^+} e^{\mu_+}_{(a)} n^{\nu_+}
\end{align}
is the three-force on the material making up the shell.  (The corresponding term for the interior is absent because the interior is flat.)  By direct calculation we find
\begin{align}\label{F}
    F_\tau = \frac{1}{4\pi R^2} \dot{V}^2\left( M'(V) - \frac{Q(V)}{R} Q'(V) \right)
\end{align}
as well as $F_A=0$.

Now we assume that the shell has no tangential stress-energy, $S_{AB}=0$.  In particular, it may be expressed as
\begin{align}
    S_{ab} = \sigma u_a u_b.
\end{align}
The component of \eqref{force} orthogonal to $u^a$ is the worldvolume geodesic equation and automatically satisfied given our assumptions so far.  The component along $u^a$ gives the continuity equation with source,
\begin{align}\label{continuity}
    D_a( \sigma u^a) = F^a u_a.
\end{align}

Using \eqref{F} and \eqref{m}, this equation becomes 
\begin{align}\label{mdot}
    \dot{m} = \dot{V}^2\left( M'(V) - \frac{Q(V)}{R(V)} Q'(V) \right).
\end{align}

The factor $\dot{V}$ can be interpreted as a blueshift, which in particular gets large near the Cauchy horizon.  A change in gravitational mass $M$ induces a blueshifted change of shell internal energy $m$, as expected.  Notice that the blueshift factor $\dot{V}$ appears \textit{squared}, indicating that this is not simple mass transfer.  If a particle of mass $m$ were subject to a mass flux of $-M'(v)$, it would gain mass at the rate $m'(v)=M'(v)$, and we would have a single power of $\dot{V}$, i.e., $\dot{m} = \dot{V} M'$.  

\section{Collapse without backreaction}\label{sec:collapse-without-backreaction}

In this section we specialize to RN spacetime, where $M$ and $Q$ are constant.  We assume that $m>0$ and $M>0$ to consider physical matter, but we do not place any restriction on the charge $Q$.  The shell mass $m$ is also constant by Eq.~\eqref{mdot}.  The parameter space divides into quadrants according to whether the spacetime is sub- or super-extremal ($|Q|<M$) or ($|Q|>M$) and whether the shell is unbound or bound ($m<M$ or $m>M$, respectively).  

We restrict to shells that start in a region with  $f>0$, which imposes  a minimum starting radius $R>r_b$ according to \eqref{branchrestriction}.  In the unbound case $m<M$ this is not an important restriction, but in the bound case $m>M$ there is also a \textit{maximum} radius $R \leq r_0$ according to \eqref{rangebound}.  Therefore, bound shells starting where $f>0$ must obey the consistency condition $r_b<r_0$, which is equivalent to 
\begin{align}\label{boundcond}
     m^2 +Q^2 - 2 m M < 0 \qquad\textrm{(bound shells).}
\end{align}
In the subextremal case this may also be written $r_- < m < r_+$, while in the superextremal case there are no solutions.  That is, there are no superextremal bound shells where $f>0$.

We now discuss the qualitative behavior of the shell.  We say that a shell ``bounces'' if it reaches an inner radial turning point and then starts re-expanding (either back to infinity or towards the Cauchy horizon in region $C$), and we say that a shell ``branches'' if it continues to shrink in size while approaching the Cauchy horizon in region $B$ (a behavior which is associated with $s$ flipping sign in our formulation).  Since the branching radius is always positive, no shell can reach $r=0$ before forming a Cauchy horizon.  That is, timelike singularities cannot be formed, either inside a black hole or naked to infinity.\footnote{An earlier version of this manuscript erroneously claimed that timelike singularities can be formed.  In fact, such singularities are forbidden by a general theorem guaranteeing positivity of the Hawking mass \cite{christodoulou1995self,Kehle:2022uvc,Kehle:2024vyt}.  I am grateful to Christoff Kehle for pointing out the theorem and helping identify my initial error.}

We discuss the bound and unbound cases separately.  In the bound case $m>M$ (which also implies subextremal $|Q|<M$), there is only an outer turning point $R \leq r_0$, so the shell radius eventually decreases monotonically.  Noting that the branching radius is between the inner and outer horizon radii ($r_-<r_b<r_+$) in the allowed range \eqref{boundcond}, we conclude that the shell always forms a trapped region ($f<0$) in which it reaches $r_b$, and flips to the $s=-1$ branch of \eqref{Vdot}, and continues toward the Cauchy horizon in region $B$.

\begin{figure}[t]    \includegraphics[width=\linewidth]{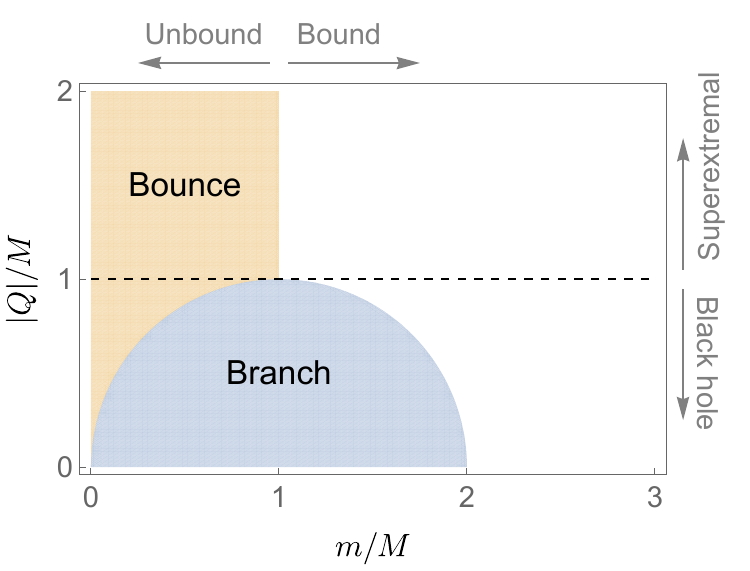}\caption{Parameter space for thin shells with a flat interior and RN exterior, assuming $m>0$ and an initial position in an $f>0$ region.  White areas are disallowed, while colored areas indicate the behavior of the shell.  Superextremal shells (top left) always bounce, meaning they re-expand to infinity.  Subextremal shells always form a black hole and Cauchy horizon.  For smaller values of $m$ they reach a minimum size and re-expand toward the Cauchy horizon (``bounce''), while for larger values of $m$ they shrink in size as they approach the Cauchy horizon (``branch'').  The bounce and branch behaviors correspond to approaching the Cauchy horizon from regions $C$ and $B$, respectively (see Fig.~\ref{fig:regions}).}\label{fig:params}
\end{figure}

In the unbound case $m<M$, the shell has an inner turning point $R \geq r_0$, which is always inside the inner horizon, $r_0<r_-$.  If $r_b>r_0$ then the shell branches and approaches the Cauchy horizon from region $B$, whereas if $r_b<r_0$ then the shell bounces and approaches the Cauchy horizon from region $C$.  The branching condition $r_b>r_0$ for $m<M$ is the same condition \eqref{boundcond} we found for the existence of bound shells.  Since all bound shells branch, we conclude that shells branch if and only if that condition is satisfied,
\begin{align}\label{branchcond}
     m^2 +Q^2 - 2 m M < 0 \qquad\textrm{(branching shells).}
\end{align}
In $m,Q$ parameter space, this defines a circle of radius $M$ centered at $m=M$ and $Q=0$.  We thus have branching behavior for all shells in this circle, bouncing behavior for unbound shells outside the circle, and no solutions for bound shells outside the circle.  These parameter regions are illustrated in Fig.~\ref{fig:params}.

\section{Collapse with backreaction}\label{sec:collapse-with-backreaction}

We now let $M(v)$ be nontrivial to mock up post-collapse accretion and/or evaporation, keeping $Q$ fixed for simplicity.  In this case the internal energy $m$ evolves as well, and we can no longer rely on effective potential reasoning.  Since $v$ now appears explicitly via $M(v)$, we find it more convenient to work with $v$ as a parameter, rather than the proper time $\tau$.  From the normalization \eqref{unorm} we have 
\begin{align}\label{Vdotkiller}
    \dot{V} = \frac{1}{\sqrt{f-2R'}},
\end{align}
which allows us to convert $\tau$-derivatives into $v$-derivatives.  In particular, Eq.~\eqref{mdot} becomes 
\begin{align}\label{mprime}
    m'= \frac{M'}{\sqrt{f-2R'}}.
\end{align}
To obtain an equation for $R'$ we express $\dot{R}^2$ in terms of $R'$ using \eqref{Vdotkiller},
\begin{align}\label{Rpquadratic}
    (R')^2 = (f-2R') \dot{R}^2,
\end{align}
which can be solved as
\begin{align}\label{Rprime}
    R' = -\dot{R}^2 \pm\sqrt{\dot{R}^4+f \dot{R}^2}.
\end{align}
Since $\dot{R}^2$ can be expressed in terms of $R$ and $m$ via \eqref{Rdot2}, we see that Eqs.~\eqref{mprime} and \eqref{Rprime} are a pair of coupled, first-order equations for $R(v)$ and $m(v)$.  In numerical simulations starting where $f>0$, the initial choice of  $\pm$ fixes the initial sign of $R'$.  As the shell and mass evolve, the $\pm$ can flip at a branching radius \eqref{rb} where $\dot{R}^2=-f$ and/or a turning radius \eqref{r0} where $\dot{R}=0$.  In fact, from \eqref{Vdot} we have $\dot{V}^{-1}=s(f+\dot{R}^2)^{1/2}-\dot{R}$ and hence
\begin{align}
    R'=-\dot{R}^2+s \dot{R}\sqrt{f+\dot{R}^2},
\end{align}
showing that the sign in \eqref{Rprime} is $\pm=s \ \textrm{sign} (\dot{R})$.  This formula can be used in numerical simulations to ensure the continuity of the solution.

However, much can be said purely analytically as long as the accretion/evaporation timescale is much slower than the formation timescale,
\begin{align}\label{slow}
    |M'(v)| \ll 1.
\end{align}
This condition is easily satisfied by the Hawking radiation of a large black hole.  It is also valid for astrophysically-reasonable accretion processes, outside of violent events like black hole mergers (where spherical symmetry would not be a good approximation anyway).

Under the assumption \eqref{slow} of slow backreaction, the RN analysis of the previous section is valid for the initial shell evolution, and we may refer to Fig.~\ref{fig:params} for its initial behavior.  The interesting range is that of black hole formation, where the shell approaches the Cauchy horizon before backreaction becomes important.  We may therefore start our analysis assuming that its deviation from $r_-$ is small,
\begin{align}\label{D}
    \mathcal{D} \equiv R-r_-, \qquad |\mathcal{D}| \ll M.
\end{align}
The metric function $f(r,v)$ \eqref{f} at the shell radius then becomes
\begin{align}\label{fnear}
    f \approx -2 \kappa_- \mathcal{D},
\end{align}
where $\kappa_-$ is the surface gravity of the inner horizon,
\begin{align}
    \kappa_- = \frac{r_+-r_-}{2 r_-^2}.
\end{align}
From \eqref{Vdotkiller} we may alternatively write $f$ as
\begin{align}\label{fexact}
    f - 2 R' = \frac{1}{\dot{V}^2}.
\end{align}
In the case without backreaction, we know that $\dot{R}$ remains finite (the shell crosses the Cauchy horizon) so we see from \eqref{Vdot} that $\dot{V}\sim f^{-1}$.  With this scaling the last term in \eqref{fexact} is negligible, implying that 
\begin{align}\label{Dprime}
    \mathcal{D}' = -\kappa_- \mathcal{D} - r_-{}'.
\end{align}
When the shell first enters the near-Cauchy-horizon regime, we may treat the surface gravity as constant and drop the last term in \eqref{Dprime}.  Thus $|\mathcal{D}|$ decreases exponentially,
\begin{align}\label{run}
\mathcal{D} \approx \mathcal{D}_0 e^{-\kappa_- v}, \qquad v \lesssim v_{e},
\end{align}
where $\kappa_-$ is the initial surface gravity and $\mathcal{D}_0$ is some arbitrary starting value for the near-inner-horizon regime.  The end of the exponential phase occurs when the two terms on the RHS of \eqref{Dprime} are of the same order of magnitude, i.e., around $v_e$ defined by
\begin{align}\label{ve}
    v_e = \frac{1}{\kappa_-}\log \frac{|\mathcal{D}_0| \kappa_-}{|r_-{}'|},
\end{align}
where the right-hand-side is evaluated at the start of the near-Cauchy-horizon regime.  

The time $v_e$ represents the time when backreaction becomes important.  Notice that this scales with the \textit{logarithm} of the accretion/evaporation rate, irrespective of its sign (see Eq.~\eqref{ve}).  The assumption of constant mass  thus breaks down on a dynamical collapse timescale, even with a small amount of accretion or evaporation.  This conclusion follows essentially from the exponential nature of the blueshift and undoubtedly holds more generally beyond our simple model.

After time $v_e$, the second term in \eqref{Dprime} becomes important.  If it has opposite sign, it arrests the descent at the critical value $\mathcal{D}=-r_-'/\kappa_-$, where things stabilize since now $\mathcal{D}'\approx 0$.  If it has the same sign as the first term, it quickly dominates, pushing $\mathcal{D}$ to flip sign and again approach the critical value.  In other words, after a time of order $v_e$ the solution is well-approximated by
\begin{align}\label{hug}
    \mathcal{D} \approx -\frac{r_-{}'}{\kappa_-}, \qquad v \gtrsim v_e.
\end{align}
That is, the shell hugs the inner horizon at a distance set by the surface gravity and timescale of accretion/evaporation.  Notice that $\mathcal{D}>0$ for accretion ($M'>0$, so that $r_-{}'<0$) and $\mathcal{D}<0$ for evaporation ($M'<0$, so that $r_-{}'>0$).  Thus the shell is in region $B$ during accretion and in region $C$ during evaporation.  Another way of stating the approximations of this section is that 
\begin{align}\label{fapprox}
    f \approx 2 R' \approx 2r_-{}'.
\end{align}

These approximations are based on $\mathcal{D}\ll 1$ and hence require the RHS of \eqref{hug} to be  small.  For non-extremal black holes this is assured by the assumption of \eqref{slow} of slow change in parameters, but the extremal limit is more delicate.  To check the consistency of the approximation, we express $r_-{}'$ as
\begin{align}\label{rmprime}
r_-{}' = M'\left(1 - \frac{M}{\sqrt{M^2-Q^2}}\right).
\end{align}
We have assumed that $M'\ll1$, but this still permits large $r_-{}'$ in the extremal limit $M\to Q$.  Furthermore, $\kappa_-$ vanishes in this limit, so $\mathcal{D}$ looks doubly large in \eqref{hug}.  However, the absorption cross-section for a black hole vanishes quickly in the extremal limit, significantly suppressing the Hawking-induced $M'(v)$.  More precisely, the near-extremal flux is  \cite{Page:2000dk}
\begin{align}\label{Mpextremal}
    M'\approx -\frac{4072}{10395\pi}\frac{\hbar(M-|Q|)^5}{M^7},
\end{align}
which easily cancels the extremal divergences in \eqref{hug} and \eqref{rmprime}.  We therefore conclude that the near-inner-horizon approximation \eqref{D} is consistent for a black hole evaporating past extremality.  

In light of this consistency, it is natural to expect that the near-inner-horizon approximation continues to hold for the entire evolution.  However, when backreaction is included we lack analytic control over the equations, and in principle, a numerical simulation could reveal a departure from the near-inner-horizon regime.  However, we see no indication of this behavior; in all simulations we consider, the shell enters the near-horizon regime and remains there for as long as we can follow the evolution.  In the case of pure evaporation ($M'(v)<0$), the results are definitive: We evolve to $m=0$ in finite time, after which the evolution can be done analytically.

Let us now analyze the behavior of $m$ after the time $v_e$ when backreaction becomes  important.  Since its evolution equation is $m'=\dot{V}M'$ \eqref{mprime} and $\dot{V}$ is becoming very large, it evolves quite dramatically.  From $m' = \dot{V} M' = (\dot{R}/R') M'$ together with \eqref{fapprox}, we may write\footnote{This approximation breaks down at turning points where $\dot{R}=R'=0$ and it was illegal to divide by $R'$.  We will only use \eqref{mnear} in the large-$m$ or small-$m$ limit, where the turning points are far from the near-horizon regime.}
\begin{align}\label{mnear}
    m' \approx \frac{|\dot{R}|_{R=r_-}}{|r_-{}'|} M'.
\end{align}
First consider the case of accretion, $M'>0$.  The RHS of \eqref{mprime} is positive, so $m$ begins to grow.  Once it is larger than all other parameters, we can get an analytic description by expanding the formula \eqref{Rdot2} for $\dot{R}$ as
\begin{align}\label{Rdotlargem}
    |\dot{R}| \approx \frac{m}{2R}, \qquad m \to \infty.
\end{align}
From \eqref{mnear} we then have
\begin{align}\label{mprimelargem}
    m' \approx \frac{M'}{|r_-'|} \frac{1}{2r_-} m.
\end{align}
Thus $m$ grows exponentially on a timescale of order the light crossing time of the black hole (and more slowly in the extremal limit---see \eqref{rmprime}).  The gravitating mass $M$ of the shell remains finite by construction, so the energetic interpretation \eqref{energy} is that extreme heating is compensated by correspondingly extreme negative gravitational energy $-m^2/(2R) \approx -m^2/(2r_-)$. We interpret this as highly blueshifted incoming radiation transforming the matter into something entirely different, with $m$ becoming an unobservable bookkeeping parameter.  

\begin{figure}
    \includegraphics[width=\linewidth]{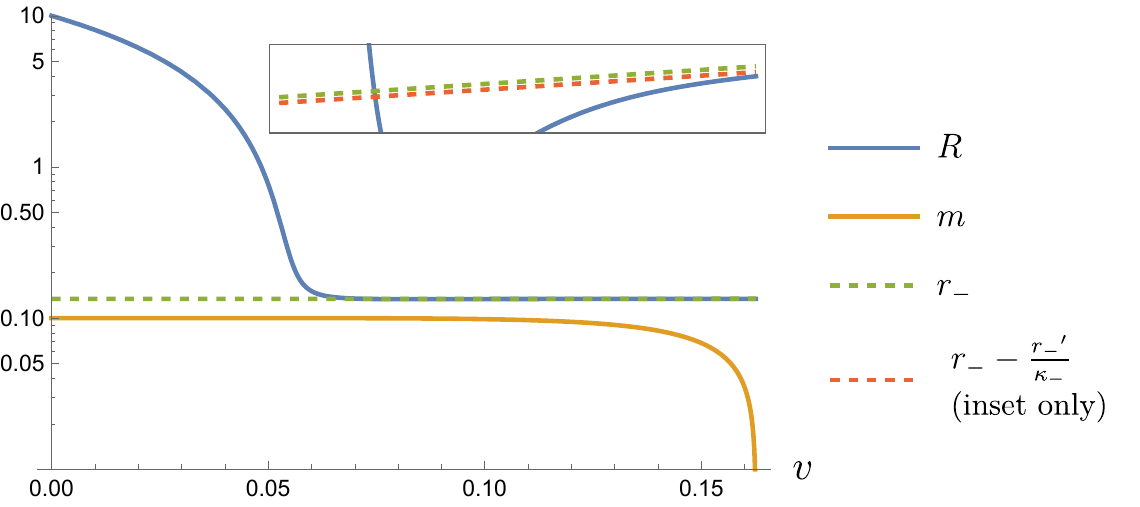}
    \caption{Numerical solution for $R(v)$ and $m(v)$ with initial values $R(0)=10$ and $m(0)=0.1$, expressed in units where the initial mass $M$ is equal to $1$.  The black hole mass evolves as $M=1-v/100$.  The shell bounces at a radius just below the inner horizon radius and then approaches the stable value \eqref{hug} set by the evaporation timescale.  At $v\approx0.16259$ the mass evolves to zero and the shell becomes null with initial radius $R\approx.13417$, just below the inner horizon at $r_-\approx.13422$.  The main plot uses a logarithmic scale, while the inset uses a linear scale showing the indicated horizontal range, together with a vertical range of $5\times 10^{-4}$.}\label{fig:bouncer}
\end{figure}

Now suppose that instead we have $M'<0$ (evaporation), so that the internal energy $m$ is decreasing.  If $m$ were already very large from a period of accretion, then it will decrease exponentially according to \eqref{mprimelargem}.  The exponential growth and decay rates depend only weakly on the black hole parameters, so $m$ will fall back down to reasonable values as long as evaporation proceeds for at least as long as accretion did.  The extreme slowness of Hawking radiation guarantees that this will happen (well) before evaporation completes.  The value of $m$ then continues to (rapidly) decrease according to the exact equation \eqref{mprime}.   When it reaches the small-$m$ regime, we can use the small-$m$ expansion of \eqref{Rdot2},
\begin{align}\label{Rdotsmallm}
    |\dot{R}| \approx \frac{2MR-Q^2}{2m R}, \qquad m \to 0.
\end{align}
The numerator is always positive in the physical range of motion, as it is the innermost accessible radius in the null limit \eqref{r0massless}.  Now \eqref{mnear} becomes
\begin{align}\label{mprimesmallm}
    m' \approx \frac{M'}{|r_-{}'|} \frac{2Mr_- - Q^2}{2r_-}\frac{1}{m}.
\end{align}
This equation takes the form $m'=-C/m$, where $C>0$ is slowly varying.  For constant $C>0$, the positive solutions are $m=\sqrt{m_0^2-2Cv}$, where $m_0>0$ is the value at $v=0$.  Thus $m$ evolves to zero in a finite time $\Delta v=m_0^2/(2C)$.  This same behavior will occur for solutions of \eqref{mprimesmallm} (where $C$ slowly varies) and hence for the exact mass evolution after sufficient evaporation.  An example is shown in Fig.~\ref{fig:bouncer}.

\begin{figure*}
    \includegraphics[width=.45\linewidth]{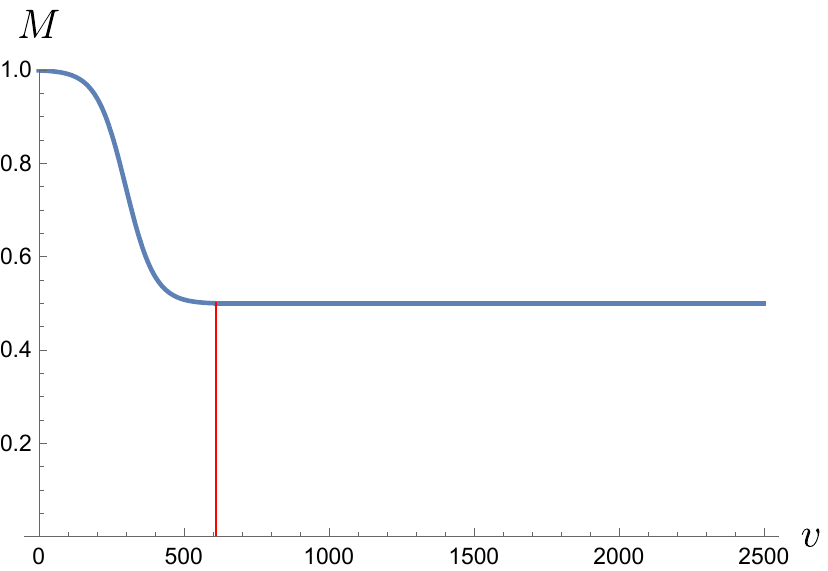} \quad \quad \quad \includegraphics[width=.45\linewidth]{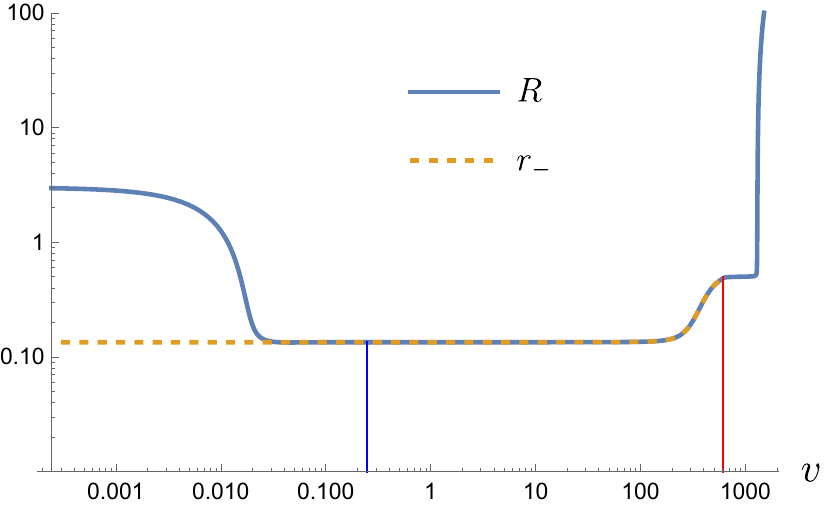}
    \caption{Evaporation past extremality with a toy mass evolution profile, shown in units of the initial black hole mass.  The charge is $Q=1/2$ and the black hole evolves from $M\approx 1$ to $M\approx 1/2$.  More precisely, the mass profile is $M=\frac{1}{2}+\frac{1}{2}( 1+e^{-c/w})/(1+e^{(v-c)/w})-\epsilon$ with $c=300$, $w=50$, and $\epsilon=10^{-3}$ until $v=611$, after which it is constant at $M=.499995$.  The time $v=611$ corresponds to the emission of the ``last quantum'' and is indicated as vertical red line.  The shell has the same parameters as Fig.~\ref{fig:bouncer}, and the moment of null conversion at $v=.249367$ is shown as a vertical blue line.  The shell trajectory is determined from Eqs.~\eqref{mprime} and \eqref{Rprime} prior to the null conversion (left of blue line) and from \eqref{exactlynull} afterwards.  The shell rapidly reaches the inner horizon radius and remains there for the entire evaporation, after which it lingers near the final horizon radius before expanding to infinity.}\label{fig:full}
\end{figure*}

We therefore conclude that the mass $m$ will reach zero exactly at some finite time $\hat{v}$, where the shell has some finite radius $\hat{R}$ and finite velocity $\hat{R}'$.  According to Eq.~\eqref{Rdotsmallm}, at this special moment we have
\begin{align}\label{pr}
    \lim_{m \to 0} m|\dot{R}|=\frac{2\hat{M}\hat{R} -\hat{Q}^2}{2\hat{R}},
\end{align}
where a hat denotes evaluation at the special time $\hat{v}$.  (We write $\hat{Q}$ for emphasis, even though we have assumed for simplicity that $Q'(v)=0$.)  Using $\dot{V}=\dot{R}/R'$, we also have 
\begin{align}\label{pv}
    \lim_{m \to 0} m|\dot{V}|=\frac{2\hat{M}\hat{R} -\hat{Q}^2}{2|\hat{R}'|\hat{R}}.
\end{align}

In order to keep $R'(v)$ finite as $|\dot{R}|\to\infty$, Eq.~\eqref{Rpquadratic} shows that $f=2R'$ \textit{exactly} (and not just approximately) at the special time $\hat{v}$.  This is the condition of a null (outgoing) worldline, and we may continue the shell evolution forward by declaring it to now be null.  Our formulation in terms of proper time breaks down, but the parameter $v$ remains useful.  The limiting value of $m |\dot{R}|$ is interpreted as the energy $m \dot{T}$ by observers inside the shell (since $\dot{T}=\sqrt{1+\dot{R}^2}$ by \eqref{Tdot}), so we will denote this quantity as $\hat{E}$,
\begin{align}
    \hat{E} = \frac{2\hat{M}\hat{R} -\hat{Q}^2}{2\hat{R}} \approx \frac{2\hat{M}\hat{r}_- -\hat{Q}^2}{2\hat{r}_-}.
\end{align}
In the second step we recall that $R$ is approximately (but not exactly) equal to the inner horizon radius.  The deviation from $r_-$ is  approximated by \eqref{hug}, and the degree of accuracy depends on how close the shell was able to approach this radius by the time of null conversion (see, for example, the inset of Fig.~\ref{fig:bouncer}).  

Observers inside the shell describe its ``total four-momentum'' $p^\mu=m u^\mu$ as 
\begin{align}
    \hat{p}^{\mu_-} = \hat{E}(1,1,0,0), 
\end{align}
i.e, they see the shell as expanding at light speed.  Observers outside the shell also consider it outgoing and null, expressed as 
\begin{align}
    \hat{p}^{\mu_+} = \frac{\hat{E}}{\hat{f}/2} (1,\hat{f}/2,0,0),
\end{align}
recalling that $R'=f/2$ exactly for the null shell, and that $f>0$ during evaporation (see  \eqref{fapprox} and discussion above).

Thus far we have described only the ``initial'' values of the null shell at the moment of transformation.  Since the shell is spherically symmetric and null, its evolution equation is trivial,\footnote{In general this argument is too glib for the future evolution of a null shell, since the vanishing of the four-momentum can signal a turning point \cite{ori1991charged}.  However, the physical range of a massless shell is $R \geq Q^2/(2M)$ (see Eq.~\eqref{r0massless}); our outgoing shell is already outside this radius and will not reach any turning point.}
\begin{align}\label{exactlynull}
    R'(v) = f/2.
\end{align}
In light of the assumption \eqref{slow} of slow evaporation (together with the even-slower evaporation \eqref{Mpextremal} occurring at the end), the solution is well-approximated by $R=r_-+\mathcal{D}$ with $\mathcal{D}\ll1$ given by \eqref{hug},
\begin{align}\label{entireevolution}
    R \approx r_- - \frac{r_-{}'}{\kappa_-}.
\end{align}
This approximation remains valid until the final ``last quantum'' pushes the spacetime above extremality and the shell's outgoing null ray is no longer inside an apparent horizon.  

When the spacetime becomes superextremal we must return to the exact equation \eqref{exactlynull}, which is now simpler since the black hole parameters are constant.  The shell will eventually expand to null infinity, but it may spend a long time near the extremal horizon radius, since the spacetime is so nearly equal to an extremal black hole.  The precise time will depend on details of how the shell became massless (which null ray it was loaded on) as well as the details of the final evaporation.  However, this phenomenon of delayed re-expansion is clearly visible in a numerical toy model  (Fig.~\ref{fig:full}).

\begin{figure}
    \includegraphics[width=.5\linewidth]{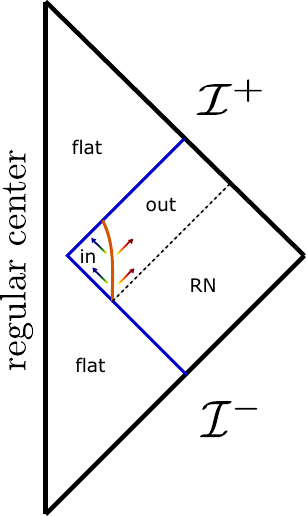}
    \caption{Penrose diagram of evaporation past extremality for a null charged shell.  The shell bounces at $r=Q^2/2M$ inside the inner horizon and then returns to infinity after evaporation.  The notation is the same as Fig.~\ref{fig:new-penrose}.}\label{fig:null-penrose}
\end{figure}

Since the shell generically becomes null, it is natural to ask what happens if it was null in the first place.  In fact, this is the simplest version of the story: An initially ingoing null shell will bounce at the null turning radius \eqref{r0massless} where its four-momentum momentarily vanishes \cite{ori1991charged}, remain outgoing during evaporation, and then return to null infinity after the apparent horizon disappears.  This simple construction is illustrated in Fig.~\ref{fig:null-penrose}.

\section{Limitations}\label{sec:limitations}

We have studied the collapse of a thin charged spherical shell including backreaction due to accretion and evaporation modeled by a charged Vaidya exterior.  We found that the black hole disappears, leaving behind an expanding null shell.  This remarkable prediction deserves a thorough critical examination, and we will now do our best.

We start with the validity of the semiclassical approximation.  Our personal viewpoint is that---in general in physics---perturbation theory should be treated as ``innocent until proven guilty''.  That is, a perturbative ansatz that satisfies consistent equations and does not predict any divergences in physical quantities should be treated as a valid prediction of the underlying equations being perturbed.   The underlying equations of quantum gravity are unknown, but for macroscopic gravitating bodies it seems clear that the relevant approximation is quantum field theory in a fixed curved spacetime.  This approximation \textit{does} predict divergences at the inner horizon, signaling strong backreaction (irrespective of the extremal limit) that has partially motivated our analysis.  However, we are unaware of any predicted divergences in the asymptotically flat exterior region that dictates the global energetics of evaporation.  Our personal bias, therefore, is to trust the semiclassical approximation even if its thermodynamic interpretation is problematic  \cite{Preskill:1991tb}.

Let us now turn to details.  One question is whether the thin shell model is too idealized.  Our intuition is that the effects of gravity and electromagnetism are so extreme that variations in material structure will not affect the outcome.  Our shell has no internal stress holding it up, so the failure to collapse can be attributed entirely to gravitational and electromagnetic forces,\footnote{Since timelike geodesics never reach the timelike singularity, intuition about a ``gravitational force'' must allow it to be repulsive at sufficiently small $r$.} which should be qualitatively similar in any reasonable model of collapse.  However, it would certainly be interesting to perform numerical studies of (e.g.) collapsing stars with incoming fluxes of energy, either positive or negative.

One may also question whether the conclusions of our model depend sensitively on the assumed spherical symmetry.  Without the symmetry there is no theorem dictating the exact geometry outside the matter, but we expect it to ``settle down'' to the Kerr-Newman (KN) metric, whose interior is similar to RN.  The ring singularity of the KN metric is causally problematic, so our suspicion is that this pathway is somehow closed, so that collapsing matter will instead approach the Cauchy horizon.  It is then quite plausible that the evolution proceeds as in our simple model, with the matter settling onto an outgoing trajectory inside the black hole as it evaporates.  If the parameter range is right for evaporation past extremality, this matter would again emerge at macroscopic size.

Another potential concern is the wild behavior of the shell internal energy $m$.  As described in Sec.~\ref{sec:collapse-with-backreaction}, it grows exponentially in response to incoming positive energy.  While there is no means of gravitationally observing the value of $m$, it does correspond to a large stress-energy and may signal departure from energy scales where the known laws of physics can be applied with confidence.  However, if the large values of $m$ are too concerning, they can be avoided by choosing a strictly negative $M'(v)<0$ and/or taking the shell to be null to begin with (Fig.~\ref{fig:null-penrose}).  In these cases the stress-energy remains small everywhere, and we still have a setup where the disappearance of a black hole can be studied without requiring a theory of quantum gravity.

The evolution to $m=0$ may also warrant concern.  Such a process would undoubtedly violate baryon number conservation, and thus is not possible with standard model physics.  However, baryon number violation is a basic feature of black hole evaporation and does not bring \textit{extra} concern for this particular model.  (Indeed, we would hope that versions of this story would eventually lead to insight about the mechanism of baryon number violation.)  On the other hand, it is slightly uncomfortable that a scenario relying on the lack of massless charged particles predicts the generation of a massless charged shell.  In response, we would say that a microscopic cutoff invisible to our model (e.g., the mass of the lightest charged particle) limits the actual nullness of the shell in a way that does not affect the predicted outcome.

One may also wonder about discharge due to Schwinger pair production.  Our scenario assumes that the electric field is small enough at the horizon for charged-particle creation  to be negligible.  Since the shell remains macroscopic (the smallest possible radius is $Q^2/2M$, the null turning point), this assumption is easily strengthened to ensure that Schwinger pair production is in fact negligible everywhere.  

One may also wonder about the Aretakis instability \cite{Aretakis:2012ei,Lucietti:2012xr,Lucietti:2012sf,Murata:2013daa,Casals:2016mel}.  Our own attitude is that this phenomenon is not so much an instability as some interesting behavior experienced by infalling observers as a perturbing field decays.  For a near-extremal black hole responding to a classical perturbation, infalling observers experience large field gradients over times of order the inverse black hole temperature, but the work done by these gradients is negligible \cite{Gralla:2016sxp}.  Furthermore, the full Aretakis behavior can be understood as regular evolution in the near-horizon geometry \cite{Lucietti:2012xr, Gralla:2017lto}, which we regard as an extremal limit co-equal in status with the usual extremal limit.  We are not concerned about any strong backreaction from this regular evolution in the extremal limit.

Another worthy concern is the well-known instability of the RN Cauchy horizon \cite{Simpson:1973ua,poisson1989inner,Poisson:1990eh,Ori:1991zz,Dafermos:2017dbw}.  We are not so concerned about the classical instability, which results in a weak singularity with small effects on incoming bodies (like the Aretakis instability).  On the other hand, the quantum version of the instability is stronger, in that energy fluxes diverge on the Cauchy horizon when backreaction is not included \cite{Hiscock:1980wr,Lanir:2018vgb,Zilberman:2019buh,Hollands:2019whz,Hollands:2020qpe,Klein:2021ctt,Zilberman:2021vgz,Zilberman:2022aum,Zilberman:2022iij,McMaken:2023tft,Klein:2023urp,Klein:2024sdd,McMaken:2024fvq,Alberti:2025mpg}.  For a shell that would approach the Cauchy horizon without backreaction, these fluxes will eventually become energetically important and modify the evolution of the shell.

Of course, this is exactly what happens in our model, which attempts to mock up these big fluxes with negative-energy null dust.  The relevant question is how faithfully our model captures the actual fluxes.  In particular, the Vaidya spacetime only has ingoing flux, whereas the actual semiclassical stress tensor has other components.  However, it is the ingoing component that gives rise to blowup on the Cauchy horizon, so it seems reasonable to ignore the other components in order to capture the largest backreaction effects.  

It is clearly important to make more direct contact with semiclassical calculations.  However, it is difficult to directly compare with calculations on a fixed black hole metric, since the Vaidya spacetime is in the opposite extreme, where the mass changes immediately as radiation is emitted.  Furthermore, existing backreaction calculations for charged black holes \cite{Diba:2002hb,Barcelo:2020mjw,Bardeen:2014uaa,McMaken:2024fvq,Boyanov:2025otp} do not include the backreaction \textit{on collapsing matter} that is essential to our proposal.  A very interesting exception is the two-dimensional model of Refs.~\cite{Frolov:2005ps,Frolov:2006is}, but this work ignores the backreaction from Hawking quanta, focusing instead on charged-particle pair creation, which is suppressed in our setup.

At the end of the day, the key question is: How realistic is the null-dust model?  All we can say at present is that ingoing negative-energy flux is a firm prediction of semiclassical calculations; that its origination just outside the horizon is one reasonable heuristic \cite{Unruh:1977ga,Chen:2017pkl,Bardeen:2014uaa,Ori:2025zhe}; and that the Vaidya/shell construction is a simple, energetically faithful way to model the process with Einstein's equations.  This first analysis is by no means definitive, but its striking implications for the nature of black hole evaporation demand more detailed investigation.

\section{Implications}\label{sec:implications}

Having duly considered a variety of objections, let us now turn to potential implications, should these ideas prove correct.

First we discuss the information paradox.  One version of the paradox is the statement that when a black hole evaporates fully and disappears, a pure state has apparently become mixed: The outgoing quanta had correlations with the ingoing quanta, but no trace of the latter remains.  
Our scenario of evaporation past extremality offers a version of the story that is much more like the proverbial ``burning lump of coal,'' where the final state is mixed in practice, but in principle one would expect to be able to recover the lost information in the coal ash that remains.  One could similarly look for correlations with the expanding remnant.

An alternative possibility is that the correlations are in fact present in the late-time vacuum, being  locally indistinguishable from vacuum fluctuations \cite{Hotta:2015yla}.  This scenario suffers from energetic difficulties when the final evaporation occurs at Planck scale, so that the final quanta have Planck energies \cite{Wald:2019ygd}.  However, for evaporation past extremality the final quanta by contrast have extremely tiny energies, and purification by vacuum entanglement seems to be an attractive possibility.

If one finds that information is preserved in evaporation past extremality, then its mechanism of preservation should still hold in the full evaporation problem.  And this sharpens the information paradox, since---unlike with the traditional conception of a singular black hole interior---we are proposing that the curvatures are sub-Planckian everywhere in spacetime until the black hole itself has shrunk to Planck size.  The belief that information is lost in black hole evaporation would then require quantum gravity to suddenly destroy all of it right at the end of evaporation.  Alternatively, one could imagine finding a mechanism of information destruction in the semiclassical physics of matter interaction with negative energy, which would relieve quantum gravity of this onerous duty.

Another flavor of the information paradox is that the black hole's thermodynamic entropy (proportional to its area) appears to become smaller than its entanglement entropy (with the outgoing Hawking radiation) after significant evaporation has occurred, at the so-called ``Page time'' \cite{Page:1993wv,Witten:2024upt}.  This is forbidden in traditional statistical mechanics and therefore suggests an alternative behavior of the entropy as evaporation proceeds.  In the scenario of evaporation past extremality we can make the evaporation arbitrarily short by starting with an arbitrarily near-extremal black hole.  That is, we can tune whether or not the Page time is reached before the end of evaporation.  It has been shown that a holographic model captures the expected turnover in entropic behavior at the Page time \cite{Penington:2019npb,Almheiri:2019psf,Almheiri:2019qdq}.  It would be very interesting to consider this scenario in the context of evaporation past extremality.

Some ideas about the information paradox suggest that information is somehow stored on, or nearby, the horizon.  This is hard to understand in semiclassical physics, where the horizon has no special local properties.  However, in our model, there is always matter inside the horizon, so it is plausible that relevant information is stored there.  In other words, the surface of the collapsing matter may act as a de facto ``holographic screen''.  Of course, the interior volume would also be available for information storage.

We emphasize that we are not proposing a resolution of the information paradox.  Our scenario suggests that expanding matter appears at the end of full evaporation, but regarding this matter as a purification suffers from all the  difficulties of the usual ``final burst'' scenario, where an arbitrarily large amount of information (accrued over the evaporation of an arbitrarily massive initial black hole) must be crammed into a finite amount of energy.  Instead, we are hoping that study of evaporation past extremality will provide insight into the full information paradox.

Information issues aside, we do wish to emphasize that this model suggests a major revision in how black hole interiors are understood.  The conventional wisdom is that the instability of the Cauchy horizon results in a ``final singularity'' that will be experienced by infalling observers in finite proper time, much like the situation for the Schwarzschild interior.  By contrast, in our scenario these observers either hit the outgoing matter inside the black hole or act themselves like the original collapsing matter, experiencing highly blueshifted incoming radiation that locks them to the inner horizon as evaporation proceeds.  This material must lose mass as evaporation proceeds, but at the final moment of black hole disappearance (be it at the near-extremal or near-zero-mass limit), what remains of it can re-emerge into the external universe.

Indeed, we predict that \textit{no singularity ever forms}.  What of the singularity theorem \cite{Penrose:1964wq}?  The theorem assumes positive energy, which Hawking radiation does not respect.  A simple physical picture emerges: the locally-small ingoing Hawking flux has amplified (blueshifted) energetic effects near the would-be Cauchy horizon, resulting in macroscopically-important energy-condition violations that prevent the formation of a singularity.  This idea likely has general validity, beyond the simple model considered here.

Can any of this story be carried over to astrophysical black holes?  We would say yes.  The Kerr-Newman interior contains a Cauchy horizon everywhere except for the measure-zero Schwarzschild solution, and the severe causal difficulties with its ring singularity suggest that the singularity cannot form dynamically.  We suspect that matter forming a Kerr-Newman black hole will approach the would-be Cauchy horizon and experience strong backreaction from incoming energy and Hawking radiation, behaving just as in our toy model.  Indeed, we would suggest that a trip inside a real black hole would reveal not a singularity, but rather an outgoing remnant of the matter that previously fell in.

\section*{Acknowledgements}
It is a pleasure to acknowledge an exceptionally stimulating workshop ``Modern Trends in
Gravity and Black Holes'' held at the University of Heraklion in September, 2025, where some of the initial ideas for this work were formulated.  I am also grateful to Vasilis Paschalidis for consultation on the numerical aspects of the project.  Finally, I am indebted to Paul Anderson and Morifumi Mizuno for extensive, illuminating discussions on this topic.  This work was supported by grants from the Simons Foundation (MPSCMPS-00001470) and the National Science Foundation (PHY-2513082).

\appendix

\bibliographystyle{utphys}
\bibliography{References.bib}
\end{document}